\documentclass[sigconf]{acmart}

\renewcommand\footnotetextcopyrightpermission[1]{} 
\AtBeginDocument{%
  }

\renewcommand\footnotetextcopyrightpermission[1]{}
\setcopyright{none}
\renewcommand\footnotetextcopyrightpermission[1]{}
\setcopyright{none}
\makeatletter
\renewcommand\@formatdoi[1]{\ignorespaces}
\makeatother
\setcopyright{none}
\copyrightyear{2018}
\acmYear{2018}
\acmDOI{}
\usepackage{makecell}

\acmConference[]{}{}{}
\acmISBN{}

\begin{document}

\title{SafePyScript: A Web-Based Solution for Machine Learning-Driven Vulnerability Detection in Python}


\author{Talaya Farasat}

\orcid{1234-5678-9012}

\affiliation{%
  \institution{University of Passau}
  \country{Germany}}
\email{tf@sec.uni-passau.de}

\author{Atiqullah Ahmadzai}
\affiliation{%
\institution{University of Passau}
    \country{Germany} }
\email{ahmadz01@ads.uni-passau.de}

\author{Aleena Elsa George}
\affiliation{%
  \institution{University of Passau}
  \country{Germany}}
\email{aeg@sec.uni-passau.de}

\author{Sayed Alisina Qaderi}
\affiliation{%
\institution{University of Passau}
    \country{Germany} }
\email{qaderi01@ads.uni-passau.de}

\author{Dusan Dordevic}
\affiliation{%
\institution{University of Passau}
    \country{Germany} }
\email{dordev01@ads.uni-passau.de}

\author{Joachim Posegga}
\affiliation{%
  \institution{University of Passau}
  \country{Germany}}
\email{jp@sec.uni-passau.de}

\renewcommand{\shortauthors}{}

\begin{abstract}
  Software vulnerabilities are a fundamental cause of cyber attacks. Effectively identifying these vulnerabilities is essential for robust cybersecurity, yet it remains a complex and challenging task. In this paper, we present SafePyScript, a machine learning-based web application designed specifically to identify vulnerabilities in Python source code. Despite Python's significance as a major programming language, there is currently no convenient and easy-to-use machine learning-based web application for detecting vulnerabilities in its source code. SafePyScript addresses this gap by providing an accessible solution for Python programmers to ensure the security of their applications.
  
  SafePyScript link: https://safepyscript.com/
\end{abstract}




\maketitle

\section{Introduction}

Finding software vulnerabilities is commonly referred to as “searching for a needle in a haystack” \cite{b11}. Identifying security vulnerabilities in software code is
highly difficult since they are rare compared to other types of software defects. For example, the infamous Heartbleed
vulnerability in OpenSSL exposed
in 2014 has affected billions of Internet users; the cause were only two missing lines of code  \cite{b43}.

Static code analyses is carried out by rule-based systems, that rely on the definition of the features of vulnerable code; unfortunately, they tend to produce a significant number of false positives \cite{b12, b13}. Manual methods of defining such features are very time-consuming and primarily depend on human experience and depth of domain knowledge. Moreover, dynamic analysis tools, particularly software penetration testing, depend on a variety of test cases generated by human experts. Consequently, they also rely heavily on the domain knowledge and expertise of security professionals to pinpoint vulnerable areas within a software system \cite{b8}.

Recognizing the constraints of these traditional methods, and with the growing accessibility of open-source software repositories, it has been recommended to adopt a data-driven approach for software vulnerability detection. Therefore, various machine learning (ML) techniques have been applied to learn vulnerable features
of source code, and to automate the process of software vulnerability identification, with varying success\cite{b1, b2, b3, b4, b5, b6, b7, b14, b16, b17, b20, b22}.

Many researchers have dedicated considerable attention to source code vulnerability detection written in different programming languages like Java, C, and C++ \cite{b1, b2, b3, b5, b8, b9, b10, b17, b21, b22}. In 2024, Python continues to maintain its prominent position as one of the top programming languages \cite{b39, b40, b41}, and also a majorly used language on GitHub \cite{b39}. Despite its popularity, Python has received comparably little attention by research on source code vulnerability detection. 

Although the studies mentioned here \cite{b6,b14,b18,b19,b20,b24} focus on the Python programming language, there is still no convenient, easy-to-use machine learning web-based application for detecting vulnerabilities in Python source code. We aim to fill this gap by developing a web-based application called SafePyScript (SafePyScript Website Link: https://safepyscript.com/), where users can easily identify vulnerabilities in Python source code. SafePyScript detects vulnerabilities using a machine learning module called BiLSTM, which is discussed in detail in our previous work \cite{b55}. Additionally, SafePyScript integrates the ChatGPT API (Turbo 3.5), enabling it to utilize the powerful capabilities of the ChatGPT model for analyzing Python source code and identifying potential software vulnerabilities. Moreover, secure Python code can also be obtained with the help of the ChatGPT model. SafePyScript provides users with comprehensive capabilities such as creating and managing user accounts, detecting vulnerabilities, achieving secure code, and downloading their reports. Additionally, users can provide feedback for future improvements.

The remaining sections of this paper are organized as follows: Section 2 highlights related work. Section 3 discusses SafePyScript in detail.  Finally, we conclude this paper with conclusion and future directions in Section 4.

\section{Related Work}
Research on source code vulnerability detection is receiving increased attention due to the escalating threat of cybercrimes. The focus of research is currently shifting from traditional static and dynamic-based models \cite{b12, b13, b15}  towards the adoption of machine learning and deep learning algorithms \cite{b1, b2, b3, b4, b5, b6, b7, b14}. Recently Sharma et al.\  \cite{b16} presented a detailed review of the studies that use machine learning techniques for source code analysis.
Similarly, Semasaba et al.\  \cite{b17} highlighted a systematic literature review (between 2014 and 2020) of software vulnerability analysis on source code using deep learning models. 

Many researchers have dedicated considerable attention to source code vulnerability detection written in different programming languages like Java, C, and C++ \cite{b5, b8, b9, b10, b17}. According to the survey of Stack Overflow in 2023, Python is the third most commonly used language, and most of the student developers use Python  \cite{b43}. Surprisingly, Python is seldomly subject to source code analysis. 

Alfadel et al.\ \cite{b19}  empirically analyze the security vulnerabilities in Python packages. Harzevili et al.\ \cite{b20} investigate the security vulnerabilities specifically in Python machine learning libraries. Ehrenberg et al.\ \cite{b24} apply named entity recognition techniques (with the help of pre-trained, BERT-based transformer model) for recognizing the Python source code vulnerabilities.

Wartschinski et al.\ \cite{b6} applied LSTM (with word2vec embeddings) for Python source code vulnerability detection. Bagheri and Hegedűs \cite{b18} presented a comparison of different text representation methods (word2vec, fastText, and BERT) with LSTM for vulnerability prediction in Python source code. Their results show that  BERT embeddings with the LSTM model achieve the best overall accuracy in predicting Python source code vulnerabilities. They do not focus on creating a state-of-the-art prediction model. Rather, they investigate the capabilities of various text embedding techniques (word2vec, fastText, and BERT) as source code representations in the context of identifying vulnerabilities in Python code. Similarly, Wang et al.\ \cite{b14} presented the empirical study of different deep learning models for Python source code vulnerability detection. They show that LSTM  and gated recurrent unit (GRU) can indeed be the best choice. Moreover, they also highlight that bi-directional LSTM and GRU with attention (using word2vec) are two optimal models for Python source code vulnerability detection. 

Following their research, in our previous work we apply BiLSTM for Python source code vulnerability detection. By leveraging the optimal settings of our BiLSTM model, we achieve a remarkable performance (average Accuracy = 98.6\%, average F-Score = 94.7\%, average Precision = 96.2\%, average Recall = 93.3\%, average ROC = 99.3\%), thereby, establishing a new benchmark for vulnerability detection in Python source code. In this work, we present SafePyScript a web tool based on BiLSTM model for vulnerability detection in Python source code. Additionally, SafePyScript integrates the ChatGPT API, enabling it to utilize the powerful capabilities of the ChatGPT model for analyzing Python code, identifying potential software vulnerabilities, and get the secure code as well. 

\section{SafePyScript}

In this section, we provide a detailed overview of SafePyScript. We begin by discussing the specific vulnerabilities that SafePyScript addresses. Next, we explain the development and architecture of SafePyScript. Finally, we provide instructions on how to use the tool effectively.

\subsection{Selection of Vulnerabilities}
 
We are examining the same software vulnerabilities highlighted in \cite{b55}, i.e., SQL Injection, Cross-Site Scripting (XSS), Command Injection, Cross-Site Request Forgery (XSRF), Path Disclosure, Remote Code Execution, and Open Redirect. Following is the brief description of each vulnerability type:

SQL Injection: An attacker inserts malicious code into SQL commands through web pages, aiming to gain unauthorized access to the database or execute commands illicitly.

Cross-site Scripting (XSS): An attacker aims to execute malicious scripts in the context of a victim's web browser by implanting malicious code to legitimate web pages. The legitimate web page or web application becomes a vehicle to deliver the malicious script to the user’s browser and to execute it there.

Command Injection: An attack in which the goal is to execute commands chosen by an atttacker on the host operating system, e.g.\ via a vulnerable application.

Cross-Site Request Forgery (XSRF): An attack that forces authenticated users to submit a request to a web application against which they are currently authenticated.

Path Disclosure: An attack where an application reveals sensitive file system information through error messages, enabling targeted attacks.

Remote Code Execution: An attacker exploits a vulnerability to execute arbitrary code on a target system remotely over a network. 

Open redirect: An attacker manipulates URLs to redirect users from a legitimate website to a malicious one.

\subsection{SafePyScript Design Architecture}

The frontend interface is built using HTML, CSS, Bootstrap, JavaScript with jQuery, and Ajax incorporated to enhance user interaction, see Figure 1. The backend is developed with the Python Django framework, and SQLite3 is used for database management. The core features are described below:
 
\textbf{User Accounts}:SafePyScript allows users to create accounts and authenticate them using a username and password to access the system.

\textbf{Machine learning Technology}: SafePyScript helps users to  detect vulnerabilities in Python source code using both the BiLSTM model and the ChatGPT model.

\textbf{Report Generation}: SafePyScript generates reports for users, which can be downloaded for further review and action.

\textbf{Secure code}: SafePyScript enables users to obtain secure code with the assistance of the ChatGPT model.

\textbf{Feedback}: SafePyScript welcomes user feedback for further improvements and modifications.

\subsubsection{SafePyScript Machine learning Technology}

SafePyScript is based on machine learning module called BiLSTM and the powerful capabilities of the ChatGPT model.

\begin{table}
	\centering
	\renewcommand{\arraystretch}{1.2}
	\begin{tabular}{|p{3.2cm}|c|c|c|c|c|c|}
		\hline
		
		\textbf{Hyper-parameter name}& \textbf{Value}  \\ 
		\hline
		Number of BiLSTM Layers (with Neurons)& \makecell{1 input layer (50), 3 hidden layers (50),\\ 1 output layer (1)} \\ \cline{1-2}
		Optimizer& Adam\\ \cline{1-2}
		Learning rate&0.001\\ \cline{1-2}
		Epoch&50\\ \cline{1-2}
		Batch size&128\\ \cline{1-2}
		Loss Function&mean\_squared\_error\\ \cline{1-2}
		Drop out rate&0.2\\ \cline{1-2}

	\end{tabular}
	\caption{Optimal BiLSTM hyper-parameters}
\end{table} 

\textbf{BiLSTM:} Bidirectional Long Short-Term Memory (BiLSTM) networks, in contrast to standard LSTM networks, process sequential data in both forward and backward directions using two separate hidden layers. These two hidden layers are connected to the same output layer, allowing the model to traverse the input data twice and capture more context. Selecting the appropriate hyperparameters is crucial for optimizing model performance and achieving benchmark results.

Based on our previous experiments \cite{b55}, the BiLSTM model was trained using Word2Vec embeddings, and the combination of hyperparameters detailed in Table 1 has achieved remarkable performance.

In this work, we decide to use our high-performance BiLSTM model in SafePyScript to efficiently detect vulnerabilities in Python source code.
 
\textbf{ChatGPT API:} SafePyScript also integrates with the ChatGPT API (Turbo 3.5) as well. By leveraging advanced language understanding capabilities, ChatGPT provides insightful analysis of Python code and suggests secure code based on a broader contextual understanding. We do effective prompt engineering for getting the most accurate and relevant responses from ChatGPT model.
\subsection{SafePyScript Usage}

Figure 1 illustrates the complete workflow of the application. Initially, the user logs in with their credentials (see Figure 2). After logging in, they create a new project and select the source of the code, choosing either to fetch it from a repository or to upload Python scripts directly. Next, the user selects a machine learning model for vulnerability detection, opting for either BiLSTM or ChatGPT (see Figures 3, 4, and 5). Once the model is selected, the user can generate secure code using the ChatGPT model (see Figure 5). With SafePyScript, the user also receives a detailed report on identified vulnerabilities. Finally, users can provide feedback to facilitate ongoing improvements.

\begin{figure}[!t]
\centerline{\includegraphics[width=\linewidth]{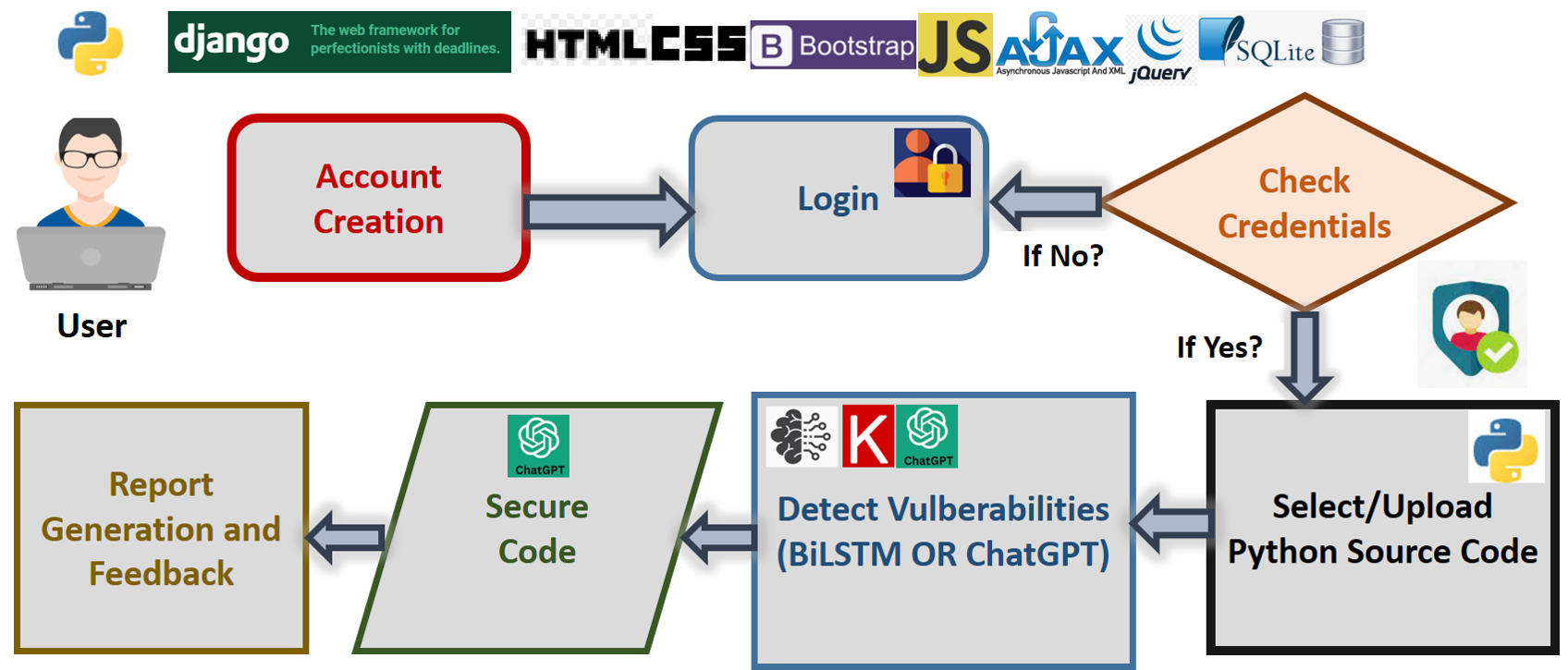}}
\caption{SafePyScript Usage}
\end{figure}

\begin{figure}[!t]
\centerline{\includegraphics[width=\linewidth]{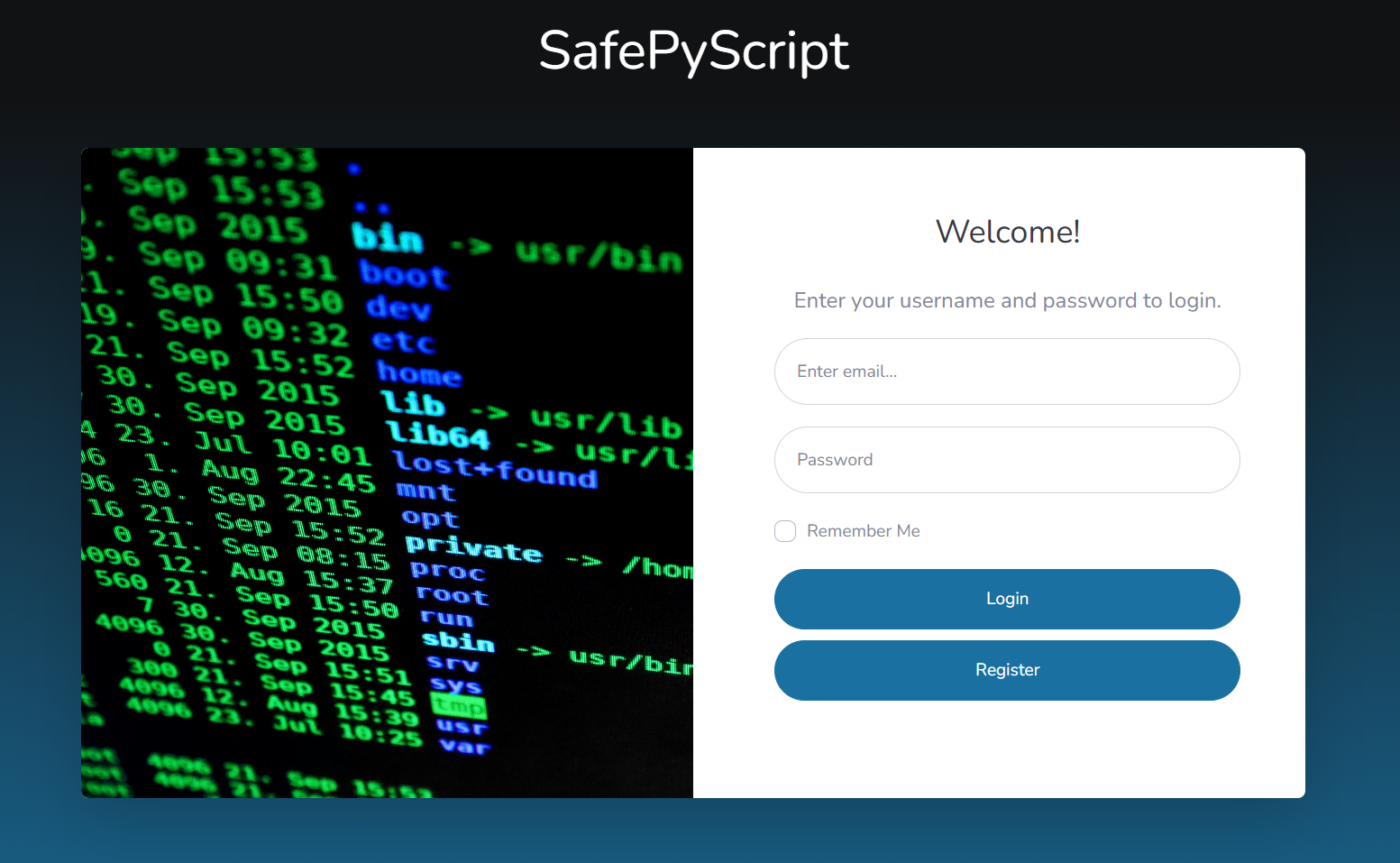}}
\caption{SafePyScript Landing Page}
\end{figure}

\begin{figure}[!t]
\centerline{\includegraphics[width=\linewidth]{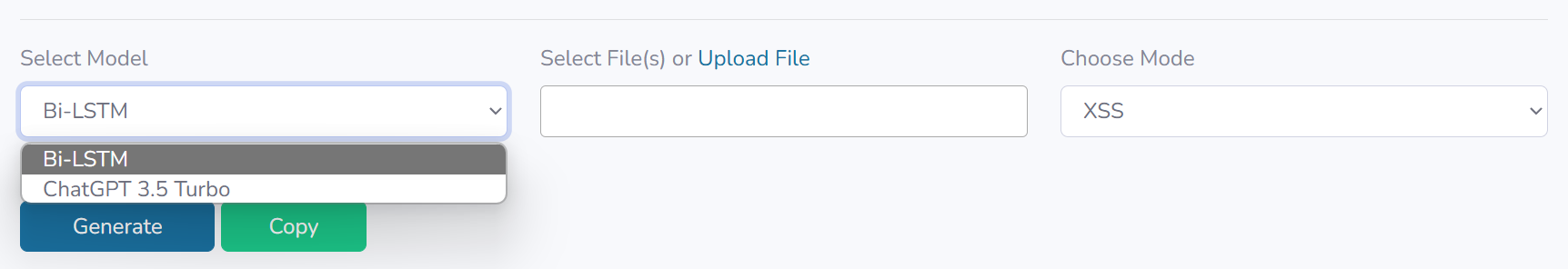}}
\caption{SafePyScript Machine learning Model Selection}
\end{figure}

\begin{figure}[!t]
\centerline{\includegraphics[width=\linewidth]{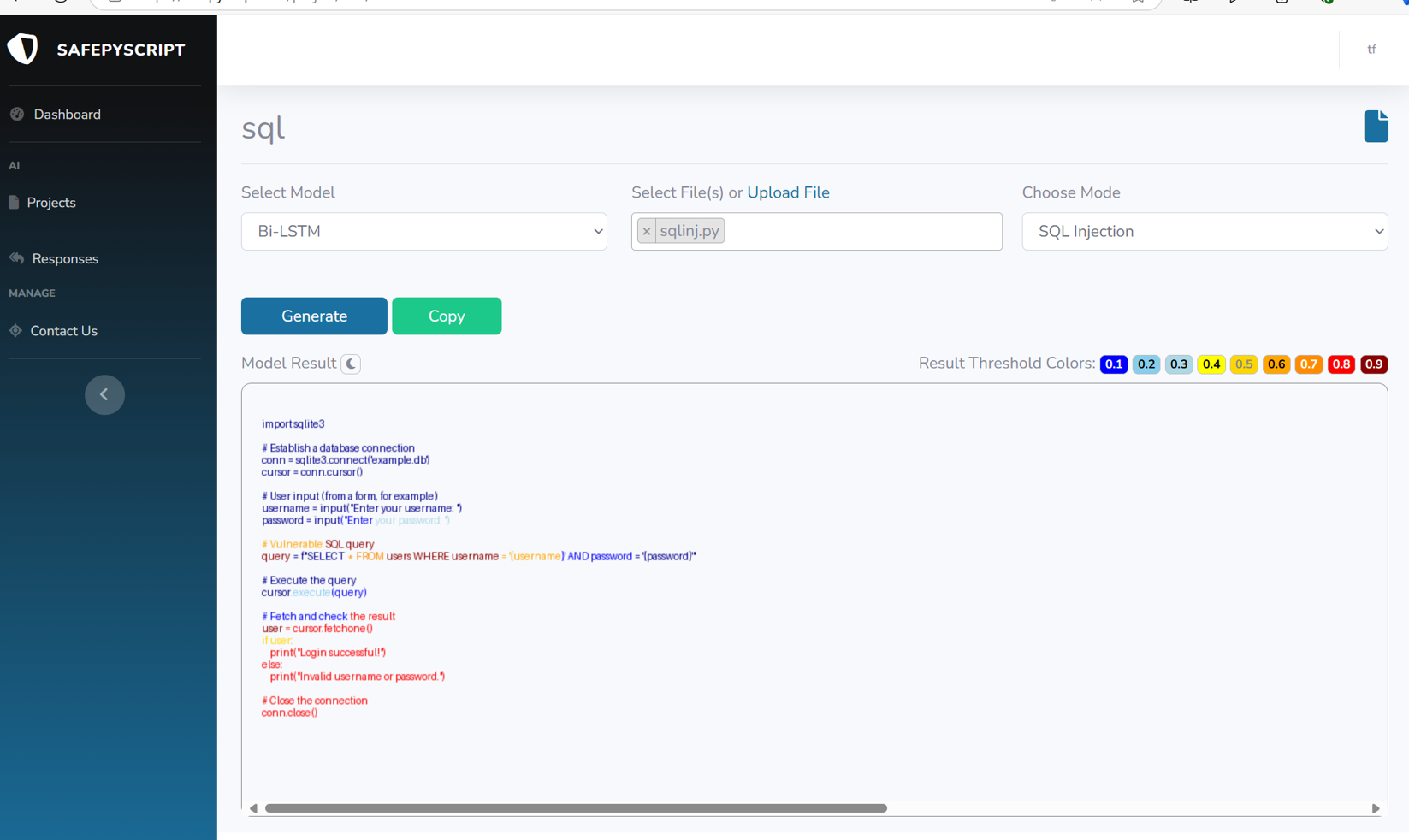}}
\caption{SafePyScript Detection of Vulnerabilty with BiLSTM Model}
\end{figure}

\begin{figure}[!t]
\centerline{\includegraphics[width=\linewidth]{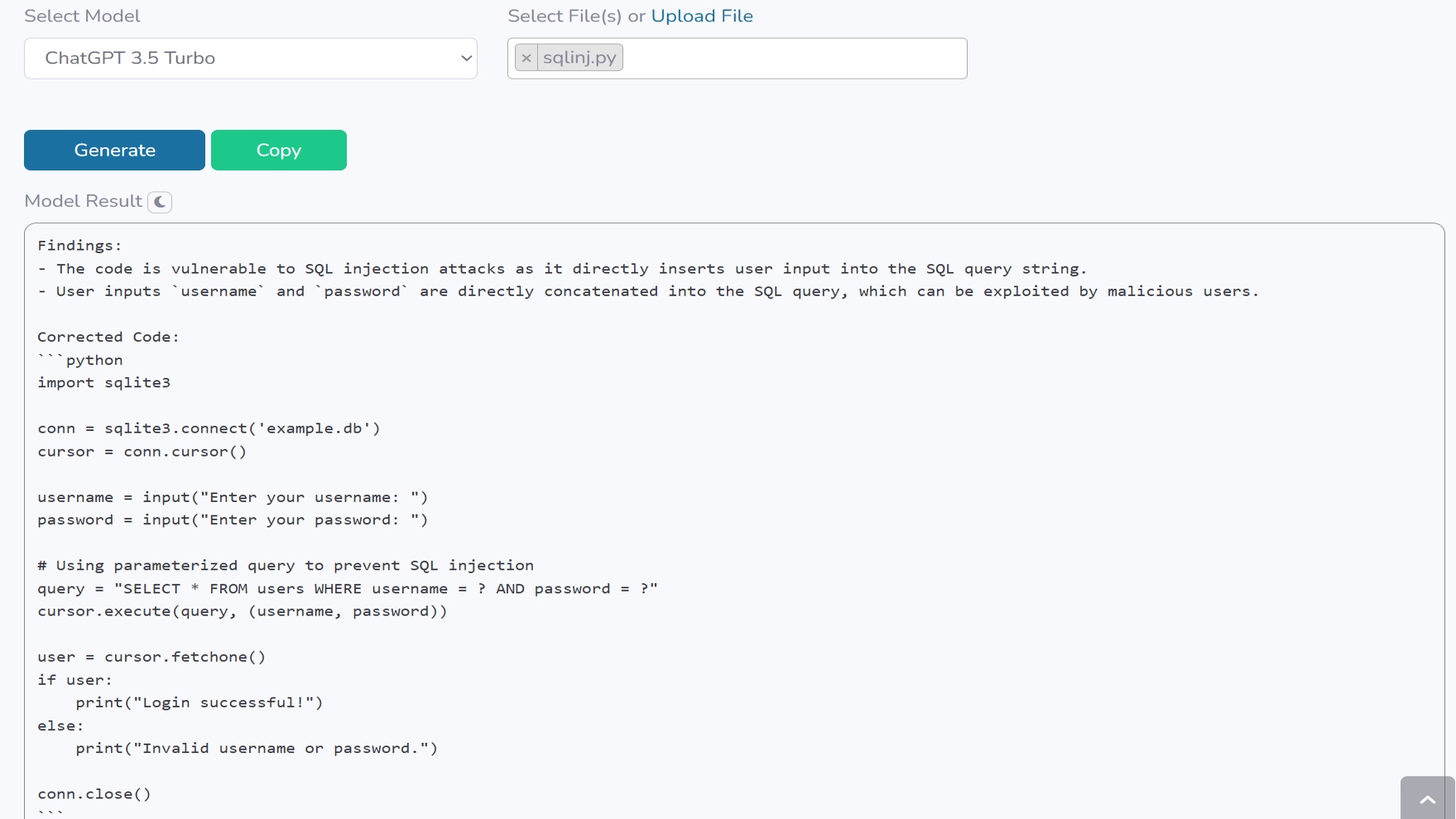}}
\caption{SafePyScript Detection of Vulnerabilty and Secure Code with ChatGPT model (Turbo 3.5)}
\end{figure}

\section{Conclusion and Future Directions}
Software vulnerabilities can lead to significant security risks, including unauthorized access, data breaches, and system disruptions. Addressing these vulnerabilities is essential for protecting sensitive information and ensuring the integrity of software applications.
In this paper, we introduced SafePyScript, a user-friendly, web-based tool designed for detecting vulnerabilities in Python source code. By integrating the high-performance BiLSTM model and the ChatGPT API, SafePyScript offers an advanced solution for comprehensive vulnerability analysis. The tool enables users to manage their profiles, detect vulnerabilities from powerful machine learning models (BiLSTM or ChatGPT), obtain reports, get secure code, and provide feedback. Despite these advancements, there are several opportunities for future improvement. Enhancing SafePyScript with additional machine learning models, could further improve detection accuracy and broaden its scope. Extending support to other programming languages and frameworks would increase its utility across different environments.  

SafePyScript Website Link: https://safepyscript.com/


\appendix

\end{document}